\documentstyle[12pt,psfig,epsf]{article} \baselineskip=2em
\def\reference{\parskip 0pt\par\noindent\hangindent 0.5 truecm}
\def\reference{\parskip 0pt\par\noindent\hangindent 0.5 truecm}

\def\emunit {cm$^{-6}$ pc}
\def\Msun  {M$_\odot$}
\def\deg   {$^\circ$}

\def\Pcos  {$\Phi^0$}

\def\Jcos  {$J_{-21}^0$}
\def\Em    {${\cal E}_m$}

\def\Ha    {${\rm H}\alpha$}

\def\etal  {{\it\ et al.}}

\begin{document}
%
%
\title{Ionized Hydrogen at Large Galactocentric Distances} \author{J.
  Bland-Hawthorn\\Anglo-Australian Observatory\\P.O. Box
  296\\Epping\\NSW 2121\\jbh@aaossz.aao.gov.au}
\date{} 
\maketitle

\begin{abstract}
  We summarize recent attempts to detect warm ionized gas at large
  galactocentric distances.  This includes searching for gas at the
  edges of spirals, in between cluster galaxies, towards extragalactic
  HI clouds, and towards high velocity clouds and the Magellanic
  Stream in the Galaxy. With the exception of extragalactic HI clouds,
  all of these experiments have proved successful. Within each class,
  we have only observed a handful of objects. It is premature to
  assess what fraction of the missing baryonic mass fraction might be
  in the form of ionized gas. But, in most cases, the detections
  provide a useful constraint on the ambient ionizing flux, and in the
  case of spiral edges, can even trace dark matter haloes out to radii
  beyond the reach of radio telescopes.
\end{abstract}


\bigskip

\section{Introduction}
The nature of `dark matter' is widely recognized as a fundamental
astrophysical problem (Carr 1994). Evidence for `missing' dark matter
exists on all scales from star clusters to the Universe itself. The
current trend is to suspect that the dark matter is mostly non-baryonic in
nature (Hegyi \& Olive 1986). But our current understanding of Big Bang
nucleosynthesis indicates that we have only observed about one tenth
of the ordinary baryonic matter in the Universe. This has encouraged
speculation that the `missing baryonic mass' is in the form of cold
gas (Pfenniger, Combes \& Martinet 1994) or hot gas (Barcons, Fabian
\& Rees 1991).  In our view, not enough attention has been paid to the
possibility that the missing mass is in the form of warm ionized gas
($\sim 10^4$K) smoothly distributed on scales much larger than the
optical disks of galaxies.  Such gas, whether photoionized or
collisionally ionized, could conceivably be detected by the
Fabry-Perot `staring' technique.  This was the original motivation for
a three year campaign with HIFI at the CFHT 3.6m telescope, and
TAURUS-2 at the AAT 3.9m and WHT 4.2m telescopes.

We now present a brief summary of results to date. It is premature
to assess the likely mass fraction concealed in ionized gas. Instead,
we find that detections of ionized gas at large galactocentric
distances have important applications to galaxy dynamics and to probing
the ambient radiation field.

\section{Intergalactic gas}

How much cold or warm debris is out there?  Searches for intergalactic
HI to date have been discouraging.  Systems like the Magellanic Stream
have only been seen in one other galaxy system (Schneider\etal\ 1983;
1989), although tidal tails have been traced out to great distances in
a number of objects (Hibbard \& Mihos 1995).  The only bona fide extragalactic
HI cloud is the HI 1225+01 SW complex (Giovanelli \& Haynes 1989); the
NE cloud is now known to have a weak optical counterpart. Other claims
have not been supported by subsequent observations (e.g. Taylor\etal\ 
1995).  The situation will no doubt improve with the Parkes Multi-Beam
and Arecibo Drift Scan surveys over the next few years in that they
are expected to find cloud masses as small as 10$^{6-7}$ \Msun\ out to
$z \leq 0.1$.

There have been various attempts to detect HI structures at large
galactocentric radius in optical emission lines. Reynolds\etal\ (1986)
achieved a marginal detection of the Leo Ring in \Ha\ but this has not
been confirmed by subsequent studies. The Magellanic Stream was
marginally detected (2.7$\sigma$) by the TAURUS-2 group
(Bland-Hawthorn, Freeman \& Quinn 1994, unpublished) in their
observations of the Sculptor group.  A problem which remains
unresolved is whether some fraction of this emission arises from
clouds within the background cluster, a controversy which began with
the original HI survey (Haynes \& Roberts 1979; Arp 1985). A more
concerted campaign has detected ionized gas in clouds MS II$-$IV along
the stream (Weiner \& Williams 1996); this result is discussed in the
Bland-Hawthorn \& Maloney (1997).  No detections in optical emission
lines have been achieved in the direction of the extragalactic HI
clouds, either towards the Haynes-Giovanelli Cloud (q.v. Vogel\etal\ 
1995) or the putative NGC 3256 cloud (McCain\etal\ 1995).  These
observations are discussed in $\S4$.

There have been claims for ionized gas from HI bridges between
galaxies, e.g. Donahue, Aldering \& Stocke (1995), using
on-band/off-band subtraction methods. It is questionable that such
methods can be used to make reliable detections at such faint flux
levels (emission measures \Em\ $\leq 1$) as they are extremely
susceptible to scattered light. Furthermore, it is important to
disperse the light spectrally to avoid any additional contribution
from the background (e.g. galaxy, night sky).  In particular, the
Reynolds Layer [NII] emission, which is transmitted by their filter,
is much brighter than the anticipated extragalactic signal.  Since the
Reynolds Layer is variable and detectable over the entire sky, it
could easily provide an excess signal when on and off-band fields are
differenced. This could explain why the claimed \Ha\ distribution
closely resembles the galaxy continuum distribution.

Is there evidence for H$^+$ clouds between galaxies? The only such
system claimed to date is the `Ant' cloud in Fornax
(Bland-Hawthorn\etal\ 1995). This was originally found from its
depolarization signature towards the huge radio lobes in NGC 1316.  We
determined that the 10 kpc cloud has a mass of roughly 10$^8$ \Msun\ 
and that it must be almost fully ionized. Follow-up HI surveys with
the VLA have revealed nothing (van Gorkom 1995, private
communication).  There are many similar depolarization `drop outs'
across the lobes which might conceivably correspond to a large
population of ionized gas clouds. Follow-up observations have been
planned to test this hypothesis.

An exciting claim was made by Williams \& Schommer (1993) that they
had detected ionized gas from a nearby absorption-line system observed
towards 3C273, even while HI had not been found in this direction.
Indeed, we were encouraged by this result in our efforts to find
intergalactic H$^+$ gas. Unfortunately, subsequent work has been
unable to confirm this claim (Bland-Hawthorn\etal\ 1994; Vogel\etal\ 
1995).

\section{What is the Fabry-Perot `staring' technique?}
To our knowledge, there are four groups actively engaged in
extragalactic work using the `staring' technique.  These programs are
based at Wisconsin (since 1971), Rutgers (since 1982), AAO/LPO/Hawaii
(since 1985), and Maryland/Caltech (since 1988). Notably, R.J.
Reynolds has made use of a 1\deg\ field to characterize the warm
ionized gas throughout the Galaxy in \Ha, [OIII], [NII] and [SII]. The
other groups typically use fields of view an order of magnitude
smaller which makes them applicable to a wider variety of
extragalactic projects than the Wisconsin system.

The `staring' technique exploits the Jacquinot advantage of the
Fabry-Perot to obtain extremely deep spectra of extended, diffuse
objects.  For a fixed gap spacing, light falling on the instrument
with wavelength $\lambda$ is dispersed according to $\lambda \propto
\cos\theta$, where $\theta$ is the off-axis field angle. The spectrum
in a narrow band is dispersed radially from the optical axis across
the field. When the data are binned azimuthally about the optical
axis, a single deep spectrum is obtained. Like long-slit
spectrometers, the instrumental profile is projected onto the detector
but the line FWHM varies across the field according to $\delta\theta\ 
\propto\ \theta^{-1}$.  At the AAT, we have already achieved H$\alpha$
emission measures of 0.2 cm$^{-6}$ pc ($2\times 10^{-19}$ erg
cm$^{-2}$ s$^{-1}$ arcsec$^{-2}$) at the 3$\sigma$ level in about 90
minutes (Bland-Hawthorn\etal\ 1995). In principle, we are able to
reach 0.02 cm$^{-6}$ pc (3$\sigma$) in about six hours using larger
optics. Due to the $\cos\theta$ factor, the raw spectrum has quadratic
sampling and needs to be resampled to a linear axis where the original
number of bins is preserved.

There are any number of potential pitfalls with the `staring'
technique. To achieve the highest sensitivity, the diffuse source
needs to be monochromatic within the detector field and have an
intrinsic line dispersion roughly equal to the instrumental profile. A
strong kinematic gradient across the field would wash out the signal
after azimuthal binning.

There is a wide class of problems relating to the atmosphere. Night
sky lines vary in brightness temporally and as a function of airmass.
Water absorption features come and go with the site humidity.
Subtracting off the background can be particularly hazardous if the
`on' and `off' spectra are obtained in separate exposures. Using data
in the same CCD frame to subtract the background has its own hazards.
Thinned chips have a bad fringeing problem at high dispersion. An
important test is to show that the spectral baseline is identical in,
say, three sectors of the field. Another weakness in many published
results to date (e.g. Songaila, Bryant \& Cowie 1989) is that the
wavelength domain of the data is typically only 5$-$10 times the
resolution element, a problem well understood by radio astronomers.
This greatly complicates defining what constitutes a reliable
continuum level as the diffuse emission we are after is usually a
faint signal situated between bright night sky lines with dominant
wings.  For this reason, we have worked hard to achieve 40\AA\ 
bandpass at \Ha\ at 1\AA\ resolution.

For systems at low velocities, we need to contend with both the
geocoronal lines {\it and} with Galactic emission. Some observers
appear to confuse night sky lines with Galactic emission (particularly
the [NII] lines), for example. It is possible to lower the background
within the resolution element of the expected signal by either using
the Earth's motion (i.e. choosing the right time of year to observe)
or by observing the object close to local midnight.

\section{Cosmic ionizing background}
The emission measure \Em\ from the surface of a cloud
embedded in a bath of ionizing radiation gives a direct gauge,
independent of distance, of the ambient radiation field beyond the
Lyman continuum edge. If we overlook problems of cloud geometry and
covering fraction, a truly extragalactic HI cloud provides a crucial
probe of the metagalactic ionizing flux.

The current best 2$\sigma$ upper limit (Vogel\etal\ 1995) for the
cosmic ionizing background is \Jcos\ $<$ 0.08 (\Pcos\ $< 3\times 10^4$
phot cm$^{-2}$ s$^{-1}$). \Jcos\ is the ionizing flux density of the
cosmic background at the Lyman limit in units of 10$^{-21}$ erg
cm$^{-2}$ s$^{-1}$ Hz$^{-1}$ sr$^{-1}$; \Pcos\ ($=2\pi$\Jcos$/h$) is the
equivalent photon flux at face of a uniform, optically thick slab.
For an electron density $n_e$ and an ion density $N_{H^+}$, 
the expected emission measure from the slab is ${\cal E}_m = \int n_e
n_{H^+}\;dl =n_e n_{H^+} L\ {\rm cm^{-6}\; pc}$ where $L$ is the
thickness of the ionized region.  The resulting emission measure for
an ionizing flux \Pcos\ is then ${\cal E}_m = 1.25\times 10^{-6}$
\Pcos\ ${\rm cm^{-6}\; pc}$.  The upper limit on \Jcos\ is sufficiently
deep that it begins to challenge various models for UV background
(Haardt \& Madau 1996).  While one would certainly want to believe
that `staring' provides a fundamental constraint, there is some
uncertainty as to whether the Giovanelli-Haynes cloud is a suitable
screen for probing the background radiation.  The HI complex appears
to be highly inclined to our line of sight (Chengalur, Giovanelli \&
Haynes 1995). What matters is the HI covering fraction {\it seen by
  the ionizing photons}, in other words, along an axis in the plane of
the sky.

More recently, we obtained (McCain\etal\ 1995) a similar 2$\sigma$
upper limit (\Jcos\ $=$ 0.09) from observations of a putative
extragalactic HI cloud in the NGC 3256 group. This was originally
`discovered' by Jayanne English (1995) but the cloud has yet to be
confirmed by follow-up observations. We are wary of Sancisi's dictum
that `most extragalactic HI clouds are not confirmed in subsequent
observations'.

\section{The ionized edges of spiral galaxies}
Sunyaev (1969) first argued that the HI disks of spiral galaxies
truncate at a few times the optical diameter because the
exponentially declining HI column eventually becomes fully ionized by
the metagalactic radiation field (see also Bochkarev \& Sunyaev 1975).
It was realized by Bland-Hawthorn\etal\ (1994) that the predicted
levels of emission (Maloney 1993; Dove \& Shull 1994) could be reached
with the `staring' technique. However, there may be  alternative
explanations in specific cases. In particular, Bland-Hawthorn \& Maloney
(1997) have shown that the young disk population could be responsible since
almost all $-$ if not all $-$ spirals have at least some warp
(Kuijken 1996; Carignan 1995, personal communication). 

We can summarize possible sources of truncation in galaxies as
follows. Space does not allow for a discussion of each case.  The
deprojected shape of the HI disk at a fixed column density is shown in
the first column.  The source of the truncation and the expected trend
in \Em\ (with increasing radius) are shown in the second and third
columns respectively:

\bigskip
\begin{tabbing}
{\it shape of HI disk}\hspace{0.5in} \= {\it source of truncation}\hspace{1.5in}
\= {\it trend with increasing radius} \\ \\
axisymmetric      \> ambient radiation field                \> \Em\ decreasing \\
bisymmetric       \> tides due to external galaxy           \> \Em\ $=$ 0 \\
mirror symmetric  \> integral-sign warp $+$ internal source \> \Em\ increasing \\
asymmetric        \> ram pressure from external medium      \> \Em\ complex \\
\end{tabbing}

In order to explore these cases, we are currently studying a range of
galaxy types in different environments.  The galaxies chosen for
detailed study include M31, M33, M83, NGC 628, NGC 3198, NGC 5266,
Fourcade-Figueroa, and the Sculptor Group.  The most detailed work has
concentrated on NGC 253 (Bland-Hawthorn, Freeman \& Quinn 1996), M33
and NGC 3198 (Bland-Hawthorn, Veilleux \& Carignan 1996).  In the case
of NGC 253, we see ionized gas with \Em\ $\sim$ 0.1 \emunit\ at and
beyond the HI disk (Fig. 1). The [NII]$\lambda 6548$/\Ha\ ratio
appears to close to unity.  This has enabled us to extend the rotation
curve by 25\% over the HI data. It is unlikely that these ratios can
be explained by a quasar-produced ionizing background. Such ratios
need additional heating without further ionization, e.g., from ram
pressure heating as the disk moves through an external medium. For
M33, we have succeeded in detecting \Ha\ at the HI edge at similar
flux levels. The WHT/CFHT observational setups did not allow us to
simultaneously observe both \Ha\ and [NII]. This galaxy has a very
substantial HI warp and the outer regions may be seeing the central
disk population (Patel \& Wilson 1995).

\section{Further Experiments}

The edges of spirals are of intrinsic interest for a number of related
experiments. As is well known, the MACHO experiment has shown that
maybe as much as half the dark matter out to the distance of the LMC
is made up of baryons. One possibility is low luminosity white dwarfs.
In his book, Sciama (1996) demonstrates that tau neutrinos with masses
24 eV and lifetimes $\sim$ 10$^{23}$ sec explain a gamut of disjoint
observations.  Sciama neutrinos share an important similarity with
white dwarfs as both could, in principle, make up the dark halo and
produce sufficient flux to ionize spiral edges (Sciama 1995).

Once weak line emission is detected, it is very difficult to identify
the source of ionization unambiguously. Thus, we have chosen to study a
diverse class of objects in different environments. In other words, to
rule out ram pressure from an external medium, we choose galaxies in
loosely bound clusters (e.g. Ursa Major). To rule out internal sources
of ionization, we choose ellipticals with extended HI disks (e.g. NGC
5266), but even here there is an uncertain contribution from ``UV
upturn'' stars. While the early results are promising, we shall
refrain from concluding this review with some general statements.

\section*{Acknowledgments}
I am grateful to my Australian and French-Canadian colleagues for
permission to mention the spiral edge detections. I am
indebted to Oxford University for the Visiting Fellowship and for
their hospitality while much of this manuscript was drafted.  I am
deeply indebted to Prof. D.W. Sciama for animated discussions and
extensive advice, some of which was acted upon. 

\medskip 
\reference Arp, H. 1985, AJ, 90, 1012

\reference Barcons, X.R., Fabian, A.C., \& Rees, M.J. 1991, Nature,
350, 685 

\reference Bland-Hawthorn, J., Ekers, R.D., van Breugel, W.,
Koekemoer, A. \& Taylor, K. 1995, ApJ, 447, L77

\reference Bland-Hawthorn, J., Freeman, K.C. \& Quinn, P.J. 1996, ApJ, submitted

\reference Bland-Hawthorn, J. \& Maloney, P.R. 1996, ApJ, submitted

\reference Bland-Hawthorn, J., Taylor, K., Veilleux, S. \& Shopbell, P.L. 1994, ApJ, 437, L95

\reference Bland-Hawthorn, J., Veilleux, S. \& Carignan, C. 1996, in preparation
\reference Bochkarev, N.G. \& Sunyaev, R.A. 1977, Soviet Astr. 21, 542 (originally 1975, AZh, 54, 957)

\reference Carr, B.J. 1994, ARAA, 32, 531

\reference Chengalur, J.N., Giovanelli, R. \& Haynes, M.P. 1995, AJ, 109, 2415

\reference Donahue, M., Aldering, G. \& Stocke, J. 1995, ApJ, 450, 45

\reference Dove, J. \& Shull, J.M. 1994, ApJ, 423, 196

\reference English, J. 1995, PhD, Australia National University

\reference Giovanelli, R. \& Haynes, M.P. 1989, ApJ, 346, L5

\reference Haardt, F. \& Madau, P. 1996, ApJ, 461, 20

\reference Hibbard, J.E. \& Mihos, C.J. 1995, AJ, 110, 140

\reference Haynes, M.P. \& Roberts, M.S. 1979, ApJ, 227, 767

\reference Hegyi, D.J. \& Olive, K.A. 1986, ApJ, 303, 56

\reference Kuijken, K. 1996, In Dark and Visible Matter in Galaxies
\& Cosmological Implications

\reference McCain, C., Freeman, K.C., English,  J. \& Bland-Hawthorn,
J. 1995, in preparation

\reference Maloney, P.R. 1993, ApJ, 414, 41

\reference Patel, K. \& Wilson, C.D. 1995, ApJ, 451, 607

\reference Pfenniger, D., Combes, F. \& Martinet, L. 1994, A\&A, 285, 79

\reference Reynolds, R.J.\etal\ 1986, ApJ, 309, L9

\reference Reynolds, R.J. 1990, in Galactic \& Extragalactic Background Radiation, eds. S. Bowyer \& C. Leinert, (Dordrecht: Kluwer), 157

\reference Reynolds, R.J. 1992, ApJ, 392, L53

\reference Schneider, S.E., Helou, G., Salpeter, E.E. \& Terzian, Y. 1983, ApJ, 273, L1

\reference Schneider, S.E.\etal\ 1989, AJ, 97, 666  

\reference Sciama, D.W. 1995, MNRAS, 276, L1

\reference Sciama, D.W. 1996, Modern Cosmology and the Dark Matter
Problem, 2nd ed. (CUP: Cambridge)

\reference Songaila, A., Bryant,  W. \& Cowie, L.L. 1989, ApJ, 345, L71

\reference Sunyaev, R. 1969, Ap. Lett., 3, 33

\reference Taylor, C.L., Brinks, E., Grashuis, R.M. \& Skillman,
E.D. 1995, ApJS, 99, 427; {\it erratum}  ApJS, 102, 89

\reference Vogel, S.N., Weymann, R., Rauch, M. \& Hamilton, T. 1995, ApJ, 441, 162

\reference Weiner, B.J. \& Williams, T.B. 1996, AJ, 111, 1156

\reference Williams, T.B. \& Schommer, R.A. 1993, ApJ, 419, L53

\section{Figure Captions}

\nobreak{\bf Figure 1.}
The emission-line spectrum at the HI edge compared with the off-field
spectrum. The difference of these spectra is shown below.  Remarkably,
the [NII]$\lambda$6548 line has a surface brightness comparable to the
H$\alpha$ surface brightness, as compared with solar-abundance HII
regions where the ratio is an order of magnitude smaller. The
azimuthally averaged galaxy continuum underlies the spectrum and
corresponds to roughly $\mu_B =$ 24 mag arcsec$^{-2}$ below [NII] 
falling to 25 mag arcsec$^{-2}$ below H$\alpha$.

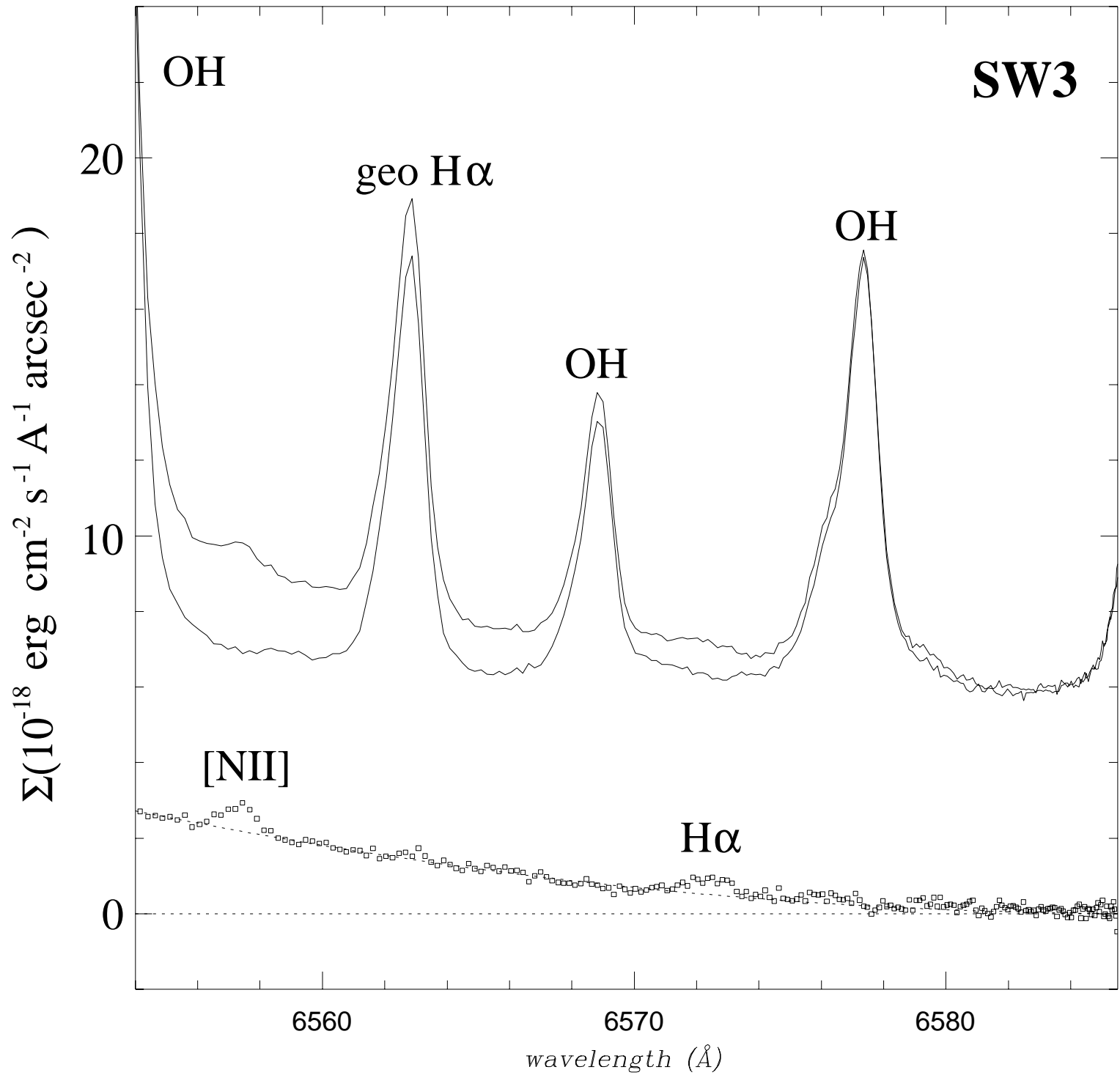
\end{document}